\documentclass[]{aastex631}

\usepackage{sidecap}
\usepackage{csquotes}
\usepackage{float}
\usepackage[many]{tcolorbox}
\usepackage{wrapfig}
\usepackage{threeparttable} 
\usepackage{booktabs}

\newcommand{\teff}{$T_{\rm{eff}}$}
\newcommand{\logg}{log($g$)}
\newcommand{\mass}{M$_\star$}
\newcommand{\z}{[Fe/H]}
\newcommand{\HI}{\ion{H}{1}}

\newcommand{\MgII}{\ion{Mg}{2}}

\newcommand{\CaII}{\ion{Ca}{2}}
\newcommand{\NaI}{\ion{Na}{1}}
\newcommand{\Lya}{Ly$\alpha$}
\newcommand{\kms}{km s$^{-1}$}

\definecolor{pink}{rgb}{1,0.1,0.6}
\definecolor{purple}{rgb}{0.4,0.01,0.8}
\definecolor{green}{rgb}{0,0.8,0}

\newcommand{\ntargets}{135}

\begin{document}

\title{Correlating Intrinsic Stellar Parameters with Mg II Self-Reversal Depths}


\author[0009-0003-5882-9663]{Anna Taylor} 
\affiliation{NASA Goddard Space Flight Center, Greenbelt, MD 20771, USA}
\affiliation{North Carolina State University, Raleigh, NC 27695-8202, USA}
\affiliation{Lunar and Planetary Laboratory, University of Arizona, Tucson, AZ 85721, USA}
\author[0009-0007-6638-5774]{Audrey Dunn}
\affiliation{NASA Goddard Space Flight Center, Greenbelt, MD 20771, USA}
\affiliation{The University of California, Los Angeles, CA 90095, USA}
\affiliation{Rochester Institute of Technology, Rochester, NY 14623, USA}
\author[0000-0002-1046-025X]{Sarah Peacock}
\affiliation{University of Maryland, Baltimore County, Baltimore, MD, 21250, USA}
\affiliation{NASA Goddard Space Flight Center, Greenbelt, MD 20771, USA}
\author[0000-0002-1176-3391]{Allison Youngblood}
\affiliation{NASA Goddard Space Flight Center, Greenbelt, MD 20771, USA}
\author[0000-0003-3786-3486]{Seth Redfield}
\affiliation{Astronomy Department and Van Vleck Observatory, Wesleyan University, Middletown, CT 06459, USA}

\begin{abstract}

The Mg II h\&k emission lines (2803, 2796 \AA) are a useful tool for understanding stellar chromospheres and transition regions due to their intrinsic brightness, relatively low interstellar medium (ISM) absorption interference, and abundance of archival spectra available. Similar to other optically thick chromospheric emission lines such as \HI\ \Lya, \MgII\ emissions commonly present with a self-reversed line core, the depth and shape of which vary from star to star. We explore the relationship between self-reversal and the stellar atmosphere by investigating the extent that fundamental stellar parameters affect self-reversal. We present a search for correlations between photospheric parameters such as effective temperature, surface gravity, and metallicity with the Mg II k self-reversal depth for a group of \ntargets\ FGKM main sequence stars with high-resolution near-ultraviolet spectra from the Hubble Space Telescope. We modeled the observed \MgII\ k line profiles to correct for ISM attenuation and recover the depth of emission line's self-reversal in relation to the intensity of the line. We used the \texttt{PHOENIX} atmosphere code to homogeneously determine the stellar parameters by computing a suite of stellar atmosphere models that include a chromosphere and transition region, and using archival photometry to guide the models of each star. We quantify the sensitivity of the visible and near-infrared photometry to chromospheric and photospheric parameters. We find weak trends between Mg II k self-reversal depth and age, rotation period, \MgII\ luminosity, temperature, and mass. All stars in our sample older than $\sim$2 Gyr or rotating slower than $\sim$10 days exhibit self-reversal. 


\end{abstract}

\section{Introduction} \label{sec:intro}

Many ultraviolet (UV) emission lines originate in the chromosphere and have the potential to serve as powerful indicators of its structure and dynamics. One indication of chromospheric properties is the self-reversal of these UV emission lines, which is the intensity dip in the center, or core, of the emission line caused by non-local thermodynamic equilibrium (non-LTE) effects. Emission lines that show self-reversal include H$\alpha$, \HI\ \Lya, \CaII\ H \& K, \NaI, and the \MgII\ h \& k doublet.  
Emission lines with self-reversals form over a wide range of altitudes in the stellar upper atmosphere. The Lorentzian line wings form deeper in the atmosphere, where conditions are close to thermal equilibrium, while the core forms at higher altitudes where the densities are low and conditions depart from thermal equilibrium. The effect is that the line intensity increases from the wings to the line core, tracing the increase in temperature with altitude, but near the very center of the line core instead of reaching a peak in intensity, the intensity declines. This traces the high altitude gas not in thermal equilibrium, where the emission efficiency is suppressed, despite the continued increase in temperature with altitude.  
The self-reversal depth is therefore intrinsically tied to properties of stellar upper atmospheres, however, it is not fully understood to what extent fundamental stellar parameters such as effective temperature (\teff), surface gravity ($log(g)$), metallicity ($[Fe/H]$), mass ($M$), radius ($R$), and magnetism might have on this property of optically thick chromospheric emission lines. Previous stellar chromosphere studies have mainly focused on line widths and intensities (e.g., \citealt{Linsky1979,Mauas2000}) with some focus on the self-reversal of the line cores of optically thick transitions \citep{Aryes1979,1992A&A...266..347P}.


\cite{Youngblood2022} found that the self-reversal depth of \ion{H}{1} \Lya\ (the brightest emission line formed in the chromosphere and transition region) correlates well with stellar surface gravity, suggesting that basic stellar structure plays a role in dictating the properties of the chromospheric emission lines. This study was limited in sample size, though, because interstellar medium (ISM) absorption contaminates the bulk of the \Lya\ line profile for the vast majority of stars, making the determination of the intrinsic \Lya\ profile and quantifying the self-reversal depth extremely difficult. The study required nearby stars with high enough radial velocities to more fully expose the line cores, which yielded a total of five GKM stars in addition to the Sun. \cite{Youngblood2022} also examined self-reversal depths of the \MgII\ k line (2796.35\AA), the brighter of the doublet, but found no apparent trends with stellar surface gravity with such a small sample size. 

Here, we expand on the \cite{Youngblood2022} study, correlating intrinsic stellar parameters and activity with the self-reversal depth of the \MgII\ k emission line. 
\MgII\ is much less affected by ISM absorption than \Lya\ owing to the lower abundance of Mg in the ISM and intrinsic narrowness of the emission line; the ISM absorption is often offset in wavelength from the stellar emission, missing it completely. There are also over 150 stars with high-resolution archival \MgII\ spectra from the Hubble Space Telescope (HST), 
allowing for more extensive analysis.

In this paper, we present correlations between Mg II self-reversal and a uniformly determined set of stellar parameters. In Section \ref{sec:obs}, we describe the stellar sample and archival spectra and photometry. Section \ref{sec:model} describes how we used \texttt{PHOENIX} \citep{HAUSCHILDT1993301, Hauschildt2006,2007A&A...468..255B} model spectra computed with empirical guidance from archival photometry to systematically determine several photospheric parameters of the stars in our sample. In Section \ref{sec:fit}, we describe how we measured the depths of the \MgII\ self-reversals. In Section \ref{sec:fit}, we present a statistical analysis of the \MgII\ self-reversal depth with various stellar parameters. In Section \ref{sec:results}, we interpret our results, and we conclude in Section \ref{sec:con}.




\section{The Stellar Sample and Archival Data}\label{sec:obs}

Drawing from the list of $>$150 stars in the HST archive with high resolution near-UV spectra (where the self-reversed cores are resolved), we created a sample of stars (Table \ref{tab:stellar_params}) limited to spectral types F, G, K, and M, luminosity classes IV, V, and VI, and detectable \MgII\ emission as determined by eye. This limits the sample to \ntargets\ stars with chromospheres without contaminating absorption signals from outflows, cool winds, or disks. In limiting the target stars to those without absorption from their winds, our sample effectively excludes low surface gravity stars (\logg $\lesssim$3.5 dex), specifically those that are young or evolved \citep{Robinson1995,Ardila2002}. We chose observations taken with the STIS E230H grating because 
the $\sim 3$ km/s spectral resolution resolves the self-reversal of the intrinsically narrow \MgII\ lines. The pipeline-reduced spectra were downloaded from the Barbara A. Mikulski Archive for Space Telescopes (MAST).The specific observations analyzed can be accessed via \dataset[DOI: 10.17909/YHRQ-T478]{https://doi.org/10.17909/YHRQ-T478}.
We combined spectra for stars with multiple orbits using a weighted average and propagated the pipeline error accordingly. In special cases where different apertures were used over the multiple orbits observed (e.g., HD9826, HD128621, and HD128620), we selected the aperture with the most orbits for coaddition. Another special case in our sample is GJ 644B, a close binary of M3.5V dwarfs (GJ 644Ba and GJ 644Bb \citep{Mazeh2001} that are spectroscopically resolved by STIS \citep{Wood2021}. In the spectra, the two stars are separated by 0.3044 \AA\ (32.6 \kms), so we separated the two emission lines at 2796.51 \AA\ for analysis. We were unable to identify which emission line belonged to which star, and arbitrarily assigned the first k line, centered at 2796.3678 \AA\, to GJ 644 Ba, and the second k line, centered at 2796.6721 \AA\, to GJ 644 Bb. The stellar parameters for these stars are very similar, therefore this arbitrary assignment does not impact our results. Note that we removed GJ 644 Bb from our sample due to a lack of available archival photometry.


We recorded literature values of each fundamental stellar parameter listed in Table \ref{tab:stellar_params} to use as a starting point for computing the stellar models in Section \ref{sec:model}. We collected minimum and maximum \teff, \logg, and \z\ parameters from SIMBAD \footnote{\url{http://simbad.u-strasbg.fr/simbad/sim-fid/}}. Stellar mass, radius, and age were found using a literature search. In instances where the stellar mass for a particular target was unknown, we used the BHAC15 models \citep{Baraffe2015} to estimate the mass based on age and \teff (if the star's age was also unknown, we assumed the star was 1 Gyr). 

\begin{figure}[t!]
    \centering
    \includegraphics[scale=0.6]{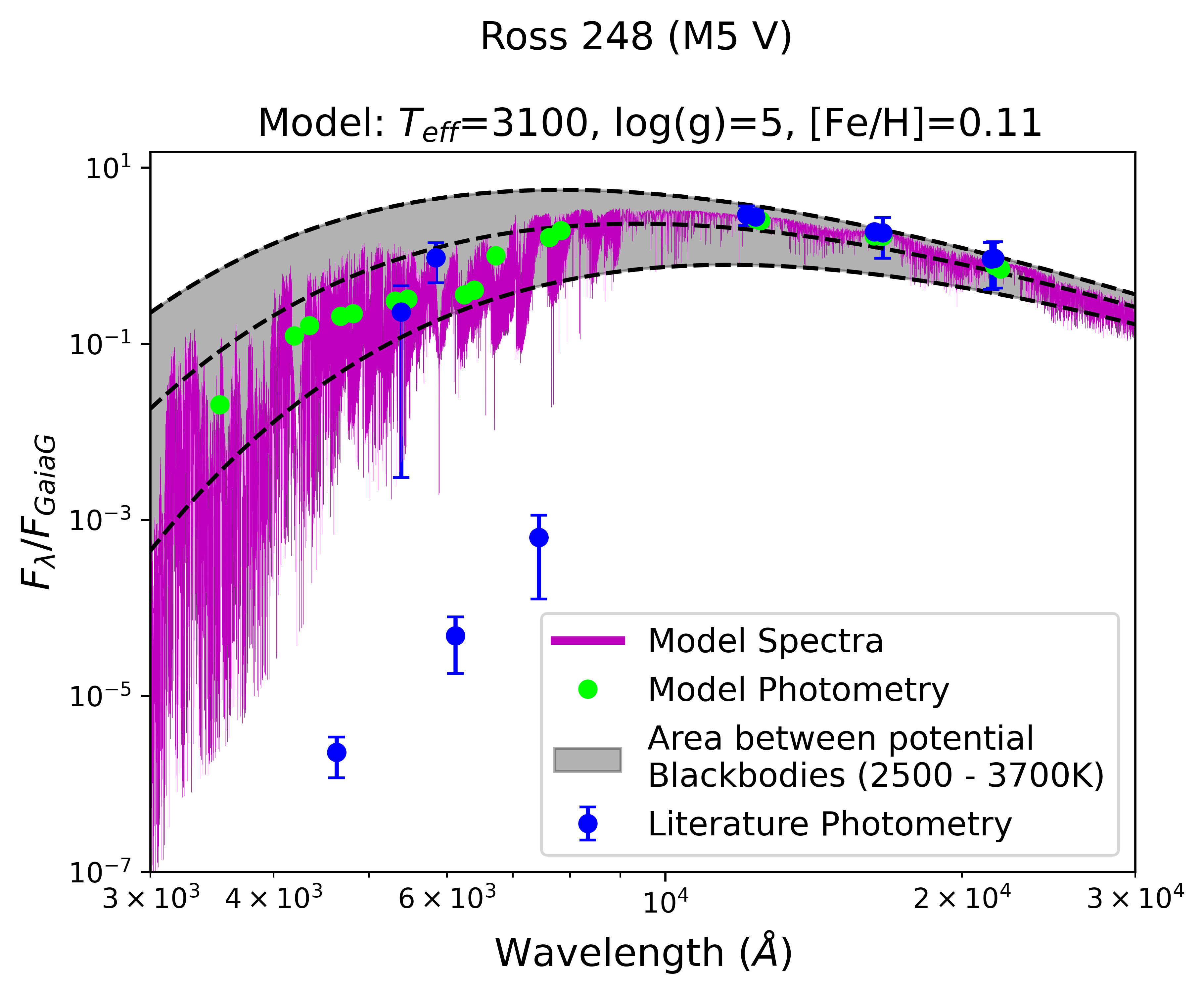}
    \caption{A model of Ross 248 (M5 V), showing the spectrum in magenta and the synthetic photometry derived from the model in green compared to the observed photometry with typical error bars from Vizier in dark blue. The area shaded in grey shows the area between \teff\ = 2500 K and \teff\ = 3700 K blackbody curves, which we use to rule out observed photometry outliers. The dashed lines represent the median and extrema blackbody curves corresponding to the range of literature values (2500-3700 K with a median value of 3100 K).}
    \label{fig:ross}
\end{figure}

We collected archival photometry in the 4200 \AA\ - 21900 \AA\ range from the VizieR Photometry Viewer \footnote{\url{http://vizier.cds.unistra.fr/vizier/sed/}} for each target star to use as empirical guidance for our homogeneous stellar parameter determination (Section \ref{sec:model}). For FGKM stars, the visible and near-infrared stellar spectrum is most sensitive to changes in \teff, \logg, and \z. In instances where there were obvious outlying photometric data points, we instituted an exclusion rule where we omitted any data points that fell outside of the range of blackbody curves for the literature-quoted minimum and maximum effective temperatures of each star. An example is shown in Figure \ref{fig:ross}, where the three data points below the bottom dashed curve were discarded.

\begin{deluxetable*}{cccc}[t]
\tablenum{1}
\tablecaption{\\Stellar Parameters for the Target Stars from the Literature}
\tablehead{Column Number & Column Name & Units & Description}
\label{tab:stellar_params}                
\startdata
    1 & Star Name & $\cdots$ & The name of the star \\
    2 & $T_{\rm min}$ & K & The minimum effective temperature \\
    3 & Reference for $T_{\rm min}$ & $\cdots$ & Reference number \\
    4 & $T_{\rm max}$ & K & The maximum effective temperature \\
    5 & Reference for $T_{\rm max}$& $\cdots$ & Reference number\\
    6 & log($g$) min & cm/sec$^2$& The minimum log(g) \\
    7 & Reference for log($g$) min & $\cdots$ & Reference number\\
    8 & log($g$) max & cm/sec$^2$& The maximum  log(g) \\
    9 & Reference for log($g$) max & $\cdots$ & Reference number\\
    10 & [Fe/H] min  & dex & The minimum  metallicity \\
    11 & Reference for [Fe/H] min & $\cdots$ & Reference number\\
    12 & [Fe/H] max  & dex & The maximum  metallicity \\
    11 & Reference for [Fe/H] max & $\cdots$ & Reference number\\
    12 & Age  & Gyr & Stellar age\\
    13 & Reference for Age & $\cdots$ & Reference number \\
    14 & Mass & M$_{\odot}$ & Stellar mass\\
    15 & Reference for Mass & $\cdots$ & Reference number\\
    16 & Radius & R$_{\odot}$ & Stellar radius \\
    17 & Reference for Radius & $\cdots$ & Reference number\\
    18 & Distance & pc & Stellar distance\\
    20 & Rotation Period & days & Stellar rotation period\\
    21 & Reference for Rotation Period & $\cdots$ & Reference number \\
\enddata
\tablecomments{This table is available in its entirety in machine-readable form. For brevity, we have numbered the references; these numbers are listed with their corresponding reference in the appendix.} 
\end{deluxetable*}

\section{Stellar Parameter Determination}\label{sec:model}

\subsection{PHOENIX Model Setup}\label{sec:modelphx}

We were motivated to generate a homogeneous set of intrinsic stellar parameters in order to minimize systematic uncertainties in our correlations between the \MgII\ self-reversal depth and fundamental stellar parameters. To consistently refine the stellar parameters for our sample with empirical guidance from visible and near-infrared photometry, we computed 1D stellar upper atmosphere models with prescriptions for the chromosphere and transition region using the multi-level non-local thermodynamic equilibrium (non-LTE) code \texttt{PHOENIX}. We used a similar setup to that described in \cite{Peacock2019}. We first computed base photosphere models in radiative–convective equilibrium corresponding to the literature value ranges of \teff, \logg, \mass, and \z\ of each star. To each of these underlying photospheres, we then superimposed a moderately increasing temperature gradient to simulate a chromosphere followed by a steep temperature gradient to simulate the transition region. The maximum temperature in the chromosphere is determined by the point at which hydrogen becomes fully ionized, occurring at temperatures near 6000 K for F and G type stars, $\sim$7000 K for K stars, and $\sim$8000 K for M stars. The criteria we use to determine when hydrogen becomes fully ionized is part of our model output. We define hydrogen as fully ionized when the abundance of neutral hydrogen drops below 10 parts per million. In our models, we set the hottest layer at the top of the transition region to be 200,000 K, since the majority of observed emission lines used in our model form at or below this temperature. The physical conditions in stellar upper atmospheres are such that radiative rates are much larger than collisional rates and radiative transfer is dominated by non-LTE effects. For our models, we perform multi-line non-LTE calculations for a small set of species that impact the spectrum in the visible and near-infrared wavelengths. We consider a total of 8,013 levels and 96,747 emission lines when computing our set of 39 atoms and ions: \ion{H}{1}, He {\small\rmfamily I -- II\relax}, C {\small\rmfamily I -- IV\relax}, N {\small\rmfamily I -- IV\relax}, O {\small\rmfamily I -- IV\relax}, Na {\small\rmfamily I -- III\relax}, Mg {\small\rmfamily I -- IV\relax}, Al {\small\rmfamily I -- IV\relax}, Si {\small\rmfamily I -- IV\relax}, Ca {\small\rmfamily I -- III\relax}, Fe {\small\rmfamily I -- VI\relax}. The consideration of this non-LTE species set most notably affects wavelengths 4500--6000 \AA\ and 9000 - 11000 \AA, with the synthetic photometry decreasing by 20--40\% in this region compared to when the model is computed assuming LTE. The addition of more non-LTE species to the models yields negligible ($<$1\%) changes to the integrated flux densities across the wavelength range corresponding to our empirical guidance (4200--21900 \AA). As a further note, we do not include modeling of dust extinction since all of our target stars lie within the Local Bubble and exhibit negligible dust extinction \citep{2003ApJ...595..858L}. 

\begin{figure}[t]
\begin{minipage}[b]{0.45\textwidth}
    \centering
    \includegraphics[scale=0.65]{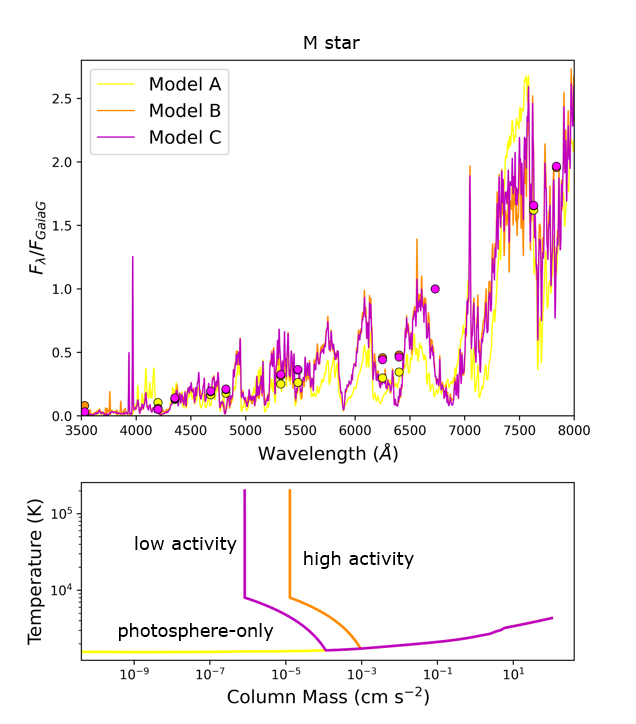}
\end{minipage}
\begin{minipage}[b]{0.50\textwidth}
    \centering
    \includegraphics[scale=0.64]{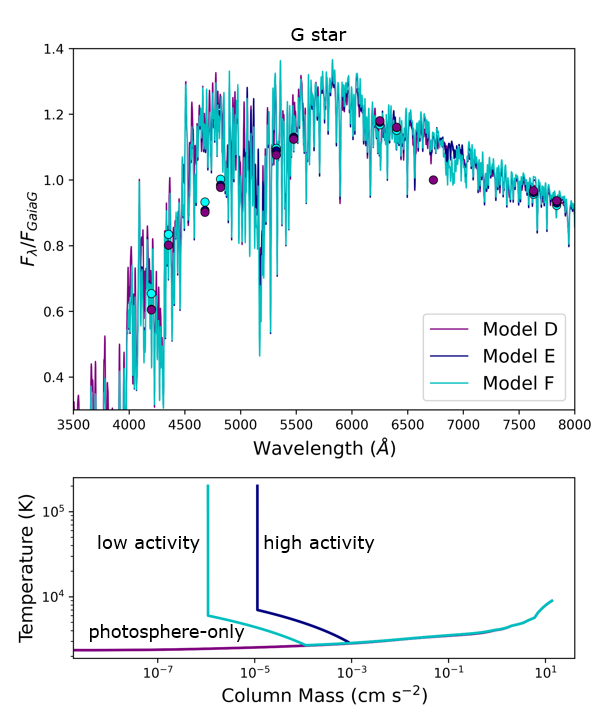}
\end{minipage}
    \caption{The upper left panel shows a comparison of three stellar spectra with \teff\ = 3000 K: Model A, computed assuming LTE (\enquote{photosphere only}), Model B, with a non-LTE, high activity chromosphere and finally, Model C, with a non-LTE, low activity chromosphere. The lower left panel shows the temperature structure with column mass corresponding in color to the three model spectra in the upper left figure. The right panel shows the same comparisons for a star with \teff\ = 5000 K. The model spectra are shown from 3500 \AA\ - 8000 \AA\ with flux density normalized to the Gaia G band. For M and G stars alike, the presence of an upper atmosphere significantly affects the visible photometry, but the differences between the high- and low-activity upper atmospheres are negligible.}
    
    \label{fig:upperatmoresponse}
    
\end{figure}

In the construction of the upper atmosphere, we altered three free parameters designating the depth at which the upper atmosphere is attached to the underlying photosphere and the thickness of both the chromosphere and transition region. The specific parameters are the column mass at the base of the chromosphere ($m_{Tmin}$), the column mass at the base of the transition region ($m_{TR}$), and the temperature gradient in the transition region ($\Delta T_{TR} = |\frac{dT}{dlogP}|$). Different states of stellar activity can be simulated by adjusting these three free parameters, which directly translates to changes in the intensities of individual chromospheric emission lines. Since our models are constrained by photometry and not spectroscopy, we investigated how changing the upper atmospheric structure affects the integrated flux density in each photometric band (Figure \ref{fig:upperatmoresponse}). We found that it is necessary to include an upper atmosphere because there is flux contribution from chromospheric emission lines in the visible bands, such as \ion{Ca}{2} H \& K (3934.77 \AA, 3969.59 \AA), \ion{Na}{1} D (5897.56 \AA, 5891.58 \AA), and H$\alpha$ (6564.62 \AA). We also found that, no matter the spectral type, large changes in the details of the upper atmosphere (high activity vs. low activity) only marginally impact the visible photometry (Table \ref{tab:percent}). These difference are negligible compared to those resulting from changing \teff\ by $\pm$50 K, log($g$) by $\pm$0.25 dex, or metallicity by $\pm$0.2 dex, which are are the smallest error bars we place on our determined stellar parameters. With this finding, we determined it was sufficient to assume a single upper atmospheric structure for each model simulating a quiescent activity state for FGKM stars: $m_{Tmin}$ = 10$^{-4}$ g/cm$^2$, $m_{TR}$ = 10$^{-6}$ g/cm$^2$, and $\Delta T_{TR}$ = $10^{8}$ K/dyne/cm$^2$. 

\newpage

\begin{deluxetable*}{c|ccccc}[t]
\centering
\tablenum{2}
\tablecaption{Change in the visible and near-infrared flux density resulting from changes in model parameters describing the upper atmosphere (\textit{middle}) versus intrinsic stellar parameters (\textit{bottom}). 
\label{tab:percent}}
\tablehead{Model & Johnson B & Johnson V & Johnson J & Johnson H & Johnson K \\
Parameter&($\lambda_{eff}$=4353 \AA)  & ($\lambda_{eff}$=5477 \AA) & ($\lambda_{eff}$=12350 \AA) & ($\lambda_{eff}$=16300 \AA) & ($\lambda_{eff}$=21900 \AA)}
\startdata
    $\nabla m_{Tmin}$ $\pm$ 100 $g/cm^2$ & $\leq$0.6\% & $\leq$0.9\% & $\leq$0.2\% & $\leq$0.1\% & $\leq$0.1\%\\
    $\nabla m_{TR}$ $\pm$  100 $g/cm^2$ & $\leq$0.6\% & $\leq$0.9\% & $\leq$0.2\% & $\leq$0.1\% & $\leq$0.1\%\\
    $\nabla \Delta T_{TR}$ $\pm$  100 K & $\leq$0.6\% & $\leq$0.9\% & $\leq$0.2\% & $\leq$0.1\% & $\leq$0.1\%\\
    \hline
    \teff\ $\pm$ 50 K &        $\sim$17\% & $\sim$12\% & $\sim$5\% & $\sim$4\% & $\sim$3\% \\
    log($g$) $\pm$ 0.25 dex  & $\sim$5\% & $\sim$3\% & $\sim$2\% & $\sim$2\% & $\sim$2\% \\
    $[$Fe/H$]$ $\pm$ 0.2 dex &         $\sim$5\% & $\sim$2\% & $\sim$2\% & $\sim$5\%  & $\sim$1\% \\
\enddata
\tablecomments{Although the precision of the archival photometry differs for every star and photometric band (shown visually in Figure \ref{fig:ross}), in general, the archival photometry cannot distinguish between the chromospheric model differences but can distinguish the temperature, surface gravity, and metallicity differences.}
\end{deluxetable*}


\newpage
\subsection{Using PHOENIX to refine intrinsic photospheric parameters}\label{sec:params}

We determine \teff, \logg, and \z\ for each star by identifying \texttt{PHOENIX} models that most closely reproduce the archival photometry. We compute small grids of models for each star, sampling around the range of literature values for each parameter in steps of 50 K for \teff, 0.25 dex for log($g$), and 0.2 dex for \z. For each model, we calculate synthetic photometry in the Johnson BVJHK, 2MASS JHK, Gaia G, HIP BT and VT, and SDSS gri bands. We use these bands because they cover extensive wavelengths sensitive to changes in the stellar parameters we are trying to refine and are available for the majority of our targets. Around $\sim 90\%$ of our target stars have observations in the Johnson BVJHK, 2MASS JHK, Gaias G, and HIP BT and VT bands. We include the SDSS gri bands because about $\sim 60\%$ of our target stars have observations in these bands, and provide additional constraints at the visible wavelengths. 

\begin{figure}[t!]
    \centering
    
        \centering
        \includegraphics[width=\textwidth]{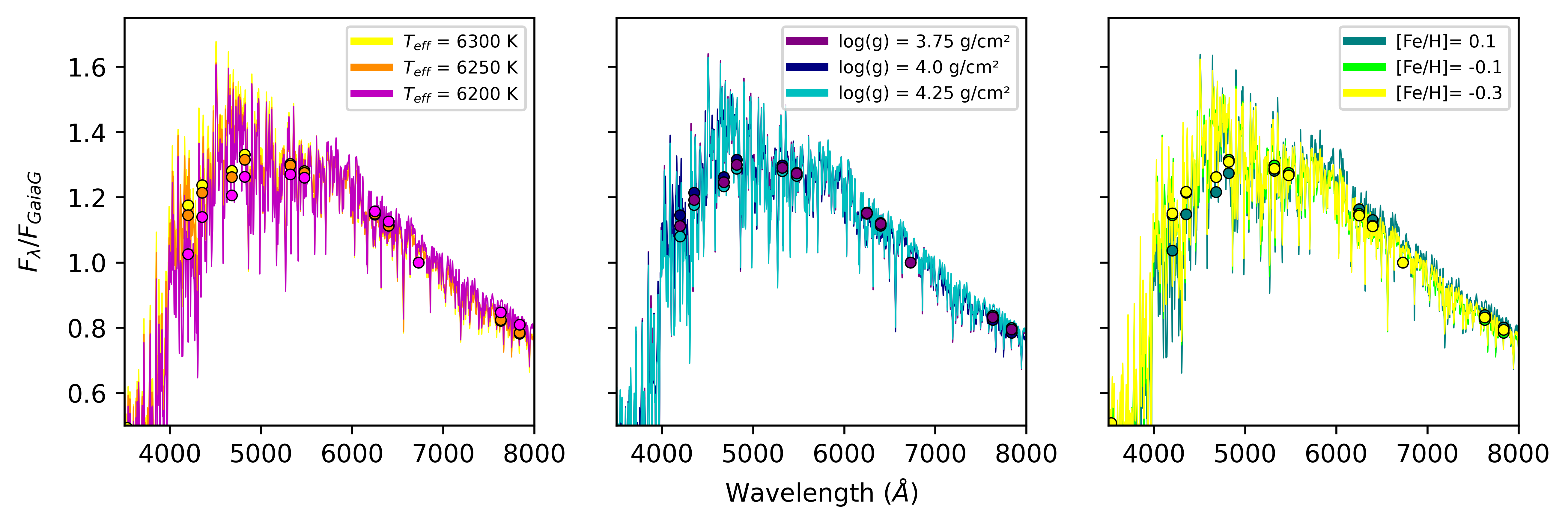}
    
    \caption{Variations of model spectra and computed photometry for the G star HD29419 (\teff = 6200 $\pm$ 50 K, log(g) = 4.0 $\pm$ 0.25 dex, and [Fe/H] = -0.1 $\pm$ 0.2 dex) normalized to the Gaia G band. The plots show the effect of changing \teff by 50 K (\textit{left}), log($g$) by 0.25 dex (\textit{middle}), and metallicity by 0.2 dex (\textit{right}). The resulting changes in synthetic photometry for each change in stellar parameter are distinguishable by the photometric measurements. } 
    
    \label{fig:photometryresponse}
\end{figure}



We show the effect that our stepped changes in \teff, \logg, and \z\ have on the flux densities in these photometric bands in Figure \ref{fig:photometryresponse}. Increasing \teff\ causes $\lambda_{max}$ to shift towards shorter wavelengths resulting in larger flux densities at visible wavelengths. Changes to \logg\ and \z\ have a less significant effect on the overall spectral shape, but do still produce measurable differences in the synthetic photometry. Increasing \logg\ results in lower emission at shorter wavelengths, while increasing \z\ results in lower flux at short wavelengths and slightly higher flux at long wavelengths.

To identify which model best represents each star, we use a reduced $\chi^2$ ($\chi^2_{\nu}$) minimization between the computed synthetic photometry and VizieR photometry. Since we are fitting for spectral shape, we normalized both the models and observations to the Gaia G band flux before calculating the reduced $\chi^2$ values. 
From each mini-grid for a given star, we identified the model with the $\chi^2_{\nu}$ value closest to 1 and adopted its \teff, \logg, and \z\ as our refined values. The typical $\chi^2_{\nu}$ value for the best-fit models ranges from 1.2--1.6. Since we are fitting for three
parameters, we chose a $\Delta\chi^2_{\nu}$ value of 3.53 to determine our model uncertainties, encompassing 68.3\% of the expected realizations \citep{Press2007}. From these models, we determined stellar radius via a distance-scaling with each star's Gaia EDR3 distance such that:
\begin{equation}
R_{\star}=\sqrt{F_{GaiaG, obs}/F_{GaiaG, mod}} \times dist^2
\end{equation}
We then calculated stellar mass from the range of radii and model-determined surface gravities for each star. Using the mass and temperature and associated uncertainties for each star, we used the Baraffe Evolutionary Model Tables \citep{Baraffe2015} to extract the approximate age, and uncertainties, of each star. In the case that the uncertainties in mass and temperature are not large enough to produce more than one age value from the Baraffe Tables, we estimate an uncertainty of 1 Gyr.
We list our refined parameters for each star in Table \ref{tab:refined}.

\begin{figure}[t!]
    \centering
    \includegraphics[scale =0.35]{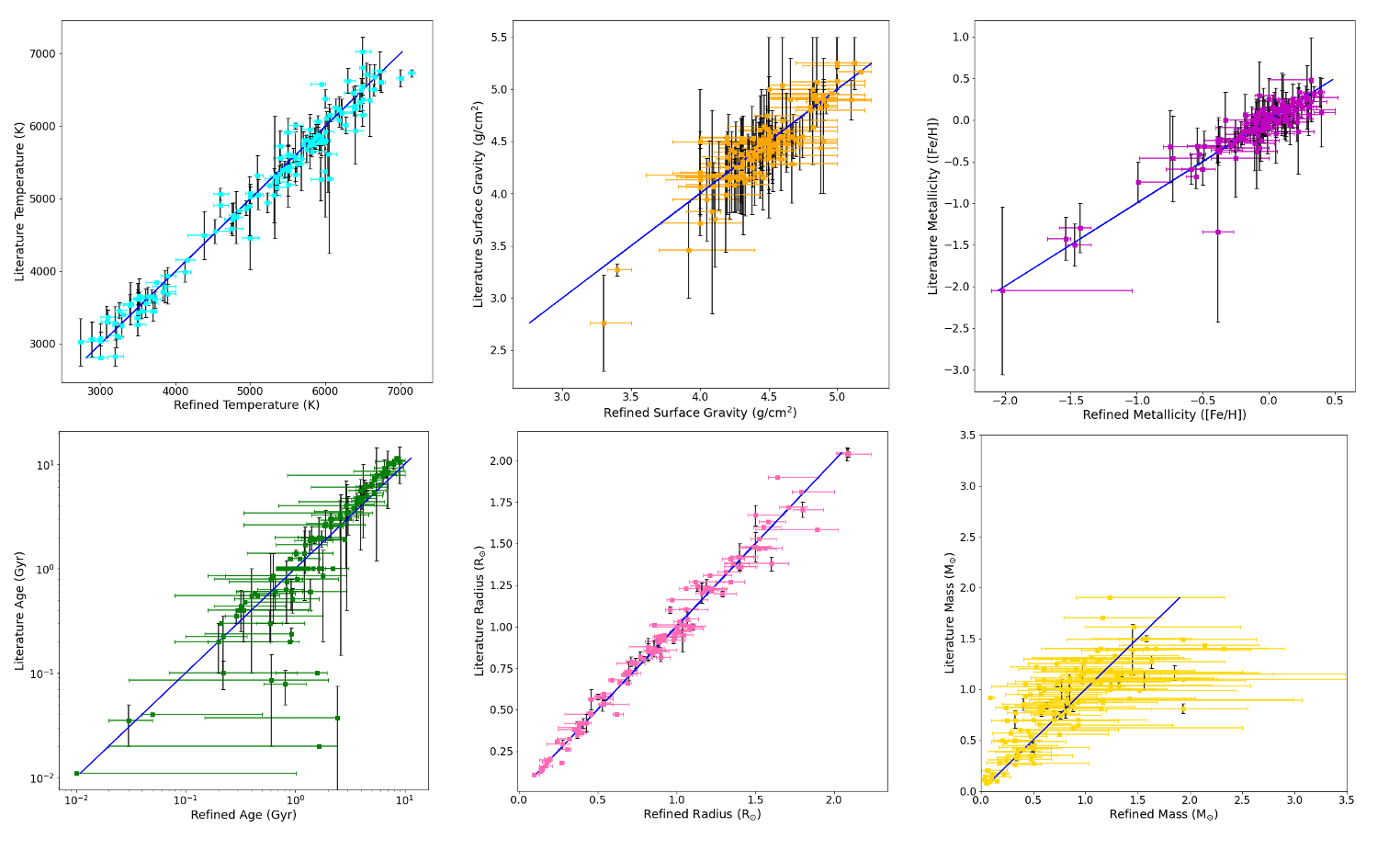}
    \caption{Literature parameters (in black) versus derived parameters (in color), temperature (\textit{top left}), surface gravity (\textit{top center}), metallicity (\textit{top right}), age (\textit{bottom left}), radius (\textit{bottom center}), and mass (\textit{bottom right}). The error bars on the literature values (black) show the range of values found in the literature for each star. The 1:1 line is shown in blue in each plot, which denotes where the parameters would fall if they matched exactly.}
    \label{fig:cross}
\end{figure}

In order to evaluate the accuracy of our derived stellar parameters, we have plotted literature parameters versus derived parameters, including the 1:1 line, which denotes where the parameters would fall if they matched exactly, in Figure \ref{fig:cross}. On the whole, our temperatures, surface gravities, metallicities, and radii agree well with literature values (with some outliers). Compared to other parameters, our mass and age values have larger uncertainties and disagree more with literature values. This is due to the coupling of the uncertainties between the parameters used to calculate mass (radius and surface gravity) and age (temperature and mass). For example for HD 199288, we find \teff = $6000^{+51}_{-65}$ K and \logg = $4.67_{-0.17}^{+0.33}$. Following Equation 1, we find radius by distance scaling with each star’s Gaia EDR3 distance and find $R_{\star}$ = $0.92_{-0.03}^{+0.06}$ $R_{\odot}$. We then combine $R_{\star}$ and \logg\ and find large uncertainties on the stellar mass, $M_{\star}$ = $1.44_{-0.53}^{+2.04}$ $M_{\odot}$. Note that the evolutionary models have a maximum age of 10 Gyr, so any literature cited value over 10 Gyr will disagree with our derived age.

\begin{table}[!htp]\centering
\tablenum{3}
\caption{Refined Stellar Parameters}

\label{tab:refined}
\scriptsize
\scriptsize
\begin{tabular}{lrrrrrrrrrrr}\toprule
Star Name &$T_{eff}$ &log($g$)  &[Fe/H] &Age &Mass & Radius &Self-Reversal &Self-Reversal & \MgII\ Flux  &Quality \\ \cmidrule{1-11}
&(K) &$(g/cm^2)$& (dex) &(Gyr) &(M$_\odot$) &($R_\odot$) &Parameter&Velocity ($km/s$)&($ergs/s/cm^2/\AA$) & \\\midrule
HD128621 &$5305_{-75}^{+95}$ &$4.5_{-0.14}^{+0.1}$ &$0.195_{-0.075}^{+0.6}$ &$5.21_{-0.96}^{+1.09}$ &$0.72_{-0.24}^{+0.43}$ &$0.89_{-0.04}^{+0.11}$ &$2.13\pm0.023$ &$0.57\pm0.051$ &$1.10 \cdot 10^{-10}$ &2 \\
& & & & & & & & & & \\
HD128620 &$5939_{-138}^{+61}$ &$4.3_{-0.5}^{+0.2}$ &$0.29_{-0.26}^{+0}$ &$3.77_{-0.37}^{+0.57}$ &$1.08_{-0.75}^{+0.66}$ &$1.22_{-0.02}^{+0.01}$ &$2.66\pm0.027$ &$1.00\pm0.048$ &$2.01 \cdot 10^{-10}$ &2 \\
& & & & & & & & & & \\
Wolf 359 &$2731_{-32}^{+140}$ &$4.6_{-0.1}^{+0.6}$ &$0.03_{-0.02}^{+0.07}$ &$0.22_{-0.12}^{+0.14}$ &$0.03_{-0.01}^{+0.1}$ &$0.14_{-0.01}^{+0.01}$ &$2.30\pm 0.11$ &$-0.19\pm0.073$ &$1.50 \cdot 10^{-13}$ & 3 \\
& & & & & & & & & & \\
HD95735 &$3500_{-100}^{+70}$ &$4.82_{-0.32}^{+0.18}$ &$-0.255_{-0.185}^{+0.225}$ &$0.72_{-0.02}^{+1.0}$ &$0.35_{-0.28}^{+0.22}$ &$0.38_{-0.13}^{+0.02}$ &$2.35\pm0.16$ &$0.21\pm0.13$ &$1.77 \cdot 10^{-13}$ &3 \\
& & & & & & & & & & \\
G 272-61A &$3000_{-190}^{+100}$ &$5.25_{-0.35}^{+0.05}$ &$0_{-0.1}^{0.1}$ &$0.7_{-0.00}^{+0.3}$ &$0.15_{-0.1}^{+0.02}$ &$0.15_{-0.02}^{+0.01}$ &$0.38\pm0.18$ &$0.00$ &$1.88 \cdot 10^{-13}$ &3 \\
& & & & & & & & & & \\
Ross 248 &$3000_{-100}^{+100}$ &$5_{-0.25}^{+0.25}$ &$0.11_{-0.2}^{+0.2}$ &$0.7_{-0.00}^{+1.08}$ &$0.10_{-0.07}^{+0.18}$ &$0.17_{-0.04}^{+0.04}$ &$2.65\pm0.19$ &$-0.065\pm0.096$ &$3.87 \cdot 10^{-14}$ &1 \\
& & & & & & & & & & \\
HD22049 &$5100_{-100}^{+63}$ &$4.5_{-0.25}^{+0.25}$ &$-0.09_{-0.2}^{+0.15}$ &$1.37_{-0.49}^{+1.57}$ &$0.56_{-0.27}^{+0.59}$ &$0.70_{-0.3}^{+0.05}$ &$1.80\pm0.071$ &$-0.34\pm0.091$ &$3.52 \cdot 10^{-11}$ &3 \\
& & & & & & & & & & \\
GJ887 &$3680_{-122}^{+70}$ &$4.88_{-0.38}^{+0.12}$ &$-0.185_{-0.055}^{+0.125}$ &$0.72_{-0.01}^{+1.04}$ &$0.23_{-0.13}^{+0.58}$ &$0.46_{-0.17}^{+0.01}$ &$2.37\pm0.20$ &$-0.086\pm0.22$ &$9.63 \cdot 10^{-13}$ &1 \\
& & & & & & & & & & \\
EZ Aqr &$3200_{-100}^{+100}$ &$5.10_{-0.1}^{+0.15}$ &$0.05_{-0.09}^{+0.05}$ &$0.7_{-0.00}^{+1.01}$ &$0.05_{-0.01}^{+0.06}$ &$0.10_{-0.00}^{+0.03}$ &$1.94\pm0.12$ &$-0.21\pm0.089$ &$9.69 \cdot 10^{-14}$ &1 \\
\vdots &\vdots &\vdots &\vdots &\vdots &\vdots &\vdots &\vdots &\vdots &\vdots &\vdots \\
\bottomrule
\end{tabular}
\tablecomments{This table is available in its entirety in machine-readable form.}
\end{table}

\newpage

\section{Mg II Self-Reversal Parameter Determination}\label{sec:fit}

In order to quantify the self-reversal depth of each star's spectrum, we defined and fit a model to find the self-reversal parameter, $p$, which defines the depth of the central dip in the emission line in relation to the intensity of the line. We fit the \MgII\ k line with a model of the intrinsic stellar emission (which includes $p$) and ISM absorption, following \cite{Youngblood2022} with slight modifications described below. 

We define the intrinsic stellar emission, $F_{intrinsic}$ as:

\begin{equation}
    F_{intrinsic} =F_{V,G}(V_{\rm radial}, A, FWHM_{\rm G}, FWHM_{\rm L}) \cdot  \exp{(-p \cdot F_{V,G}^{norm})} + F_{continuum}(c_0, c_1, c_2).
\end{equation}

\noindent $F_{intrinsic}$ describes how the emission line would appear if there were no ISM absorption. We use 
a Voigt profile or Gaussian profile $F_{V,G}$ (the \texttt{astropy Voigt1D} or \texttt{Gaussian1D} functions), which is a function of the radial velocity ($V_{\rm radial}$), amplitude ($A$), Gaussian full width at half maximum ($FWHM_{\rm G}$), and Lorentzian full width at half maximum ($FWHM_{\rm L}$) if using a Voigt profile, which are further defined in Table \ref{tab:MgIIFitParameters}. The self-reversal is applied as a peak-normalized Voigt or Gaussian absorption profile ($F_{V,G}^{norm}$) that mimics the shape of $F_{V,G}$ by using the same $FWHM_{\rm G}$ and $FWHM_{\rm L}$ but is centered in velocity space at $V_{\rm radial}$ + $V_{SR}$, where $V_{SR}$ is the self-reversal velocity. 
We used a Gaussian profile for only 22 stars, due to their low signal-to-noise spectra where the wings of the Voigt profile were not detectable. The continuum model, $F_{continuum}$, is a modification from \cite{Youngblood2022} describing the \MgII\ photospheric absorption surrounding the \MgII\ emission line. We used the \texttt{numpy Polynomial} function up to the second degree for $F_{continuum}$. Only seven of our warmest stars required a second-degree polynomial; all others used a constant or linear continuum, because STIS cannot detect the faint \\MgII\ photospheric absorption surrounding the \MgII\ emission  lines of cooler stars.

\begin{figure}
    \centering
    \includegraphics[width=\textwidth]{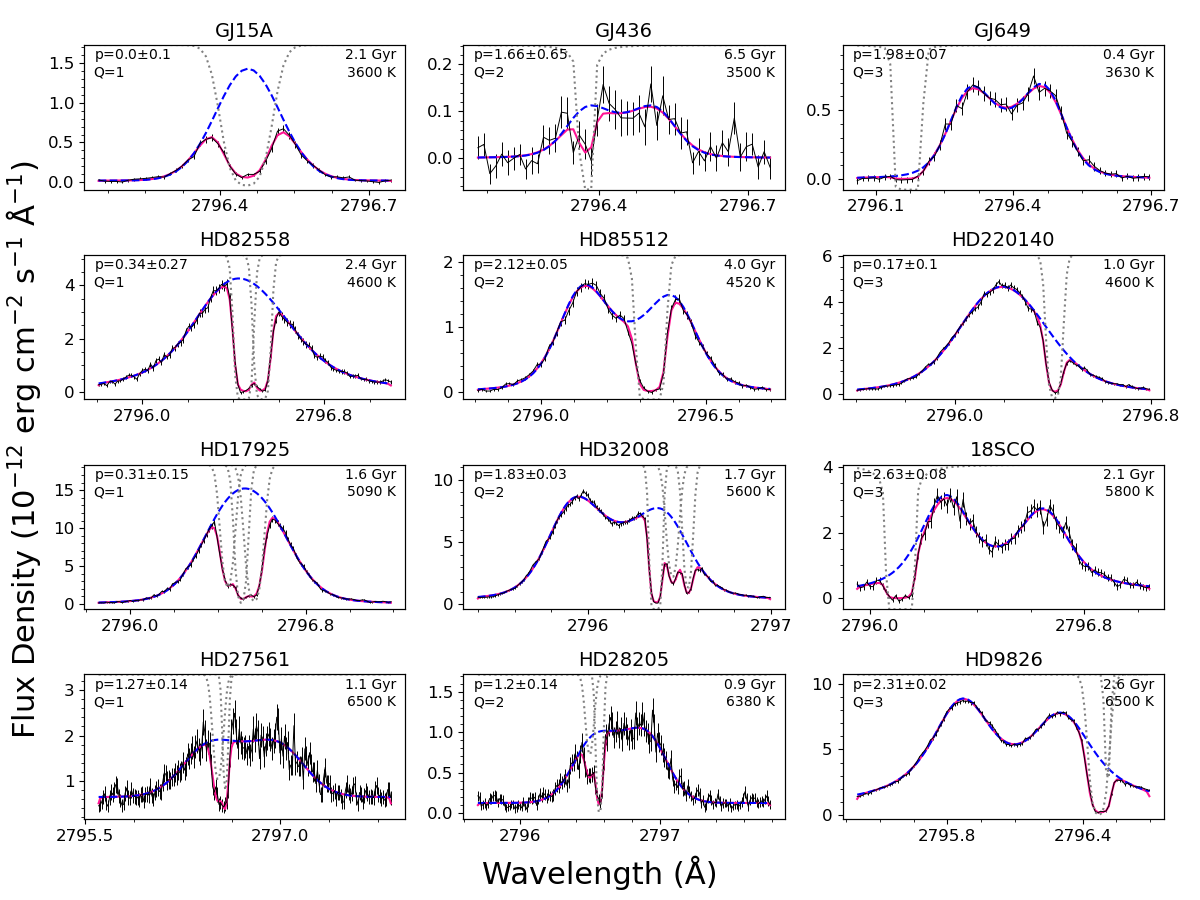}
    \caption{Stars spanning the range of observed self-reversal parameters, $p$, and quality factors, $Q$, are shown. The rows are organized by spectral type with M dwarfs in the top row and F dwarfs in the bottom row, and the columns are organized by quality factor with the lowest quality in the left column and the highest quality in the right column. In each panel, the STIS E230H data and uncertainties are shown in black, the intrinsic stellar profile in dashed blue, the interstellar absorption in dotted gray, and the best fit model in pink. The ISM absorption lines are shown with values ranging from 0 to 1 scaled to the vertical range of each panel. For visual clarity, the polynomials used to fit the stellar continuum are not shown. Self-reversal ($p$), quality factor ($Q$), stellar age, and effective temperature are printed in each panel.}
    \label{fig:BestFitEx}
\end{figure}

The self-reversal parameter, $p$, defines the depth of the central dip in the emission line in relation to the intensity of the line. Figure \ref{fig:BestFitEx} shows our model fits to twelve stars where we recovered values of $p$ from 0-2.63. 
When $p=0$, there is no self-reversal present, thus the emission line is a pure Voigt (or Gaussian) profile (e.g., Figure \ref{fig:BestFitEx}; GJ15A). When $0<p<1$, the emission line top becomes flatter as it is approaching self-reversal (e.g., Figure \ref{fig:BestFitEx}; HD82558). When $p\approx1$, a flat top emission line is produced (e.g., Figure \ref{fig:BestFitEx}; HD28205). When $p>1$, self-reversal occurs, with larger $p$ values indicating deeper central dips, (e.g., Figure \ref{fig:BestFitEx}; 18Sco).

\begin{deluxetable}{cccc}
\tablecaption{\MgII\ Observed Stellar Line Model Parameters}
\tablenum{4}
\tablehead{
 \colhead{Parameter Name} & \colhead{Variable} & \colhead{Units} & \colhead{Range} 
 }
\startdata
    \shortstack{Self-Reversal \\ Parameter} & $p$ & unitless & $0<x<4$ \\ 
    \shortstack{Emission Line \\ Radial Velocity} & $V_{\rm radial}$  & \kms & $-300<x<300$ \\ 
    \shortstack{Self-Reversal \\ Radial Velocity} & $V_{SR}$ & \kms & $-5<x<5$ \\ 
    Amplitude & $A$ & erg cm$^{-2}$ s$^{-1}$ \AA$^{-1}$ & $-16<x<-8$ \\ 
    \shortstack{Lorentzian Full Width \\ Half Maximum} & $FWHM_{\rm L}$ & \kms & $0<x<200$ \\ 
    \shortstack{Gaussian Full Width \\ Half Maximum} & $FWHM_{\rm G}$ & \kms & $0<x<200$ \\ 
    ISM \MgII\ Column Density & $N$(\MgII) & cm$^{-2}$ & $10^{8}<x<10^{17}$ \\ 
    \shortstack{ISM \MgII\ Doppler \\ Broadening Parameter} & $b_{MgII}$ & \kms & $0.5<x<7$ \\ 
    ISM \MgII\ Radial Velocity & $V_{MgII}$ & \kms & $-50<x<50$ \\ 
    Continuum Constant & $c_0$ & erg cm$^{-2}$ s$^{-1}$ \AA$^{-1}$ & $-10^{-10}<x<10^{-10}$ \\ 
    Linear Continuum Coefficient & $c_1$ & erg cm$^{-2}$ s$^{-1}$ \AA$^{-1}$ & $-10^{-10}<x<10^{-10}$ \\ 
    Quadratic Continuum Coefficient & $c_2$ & erg cm$^{-2}$ s$^{-1}$ \AA$^{-1}$ & $-10^{-10}<x<10^{-10}$ \\ 
\enddata
\tablecomments{We adopted rest wavelength $\lambda_0$ = 2796.3543 \AA\ for the \MgII\ k line \citep{Morton2003}. The Range column denotes the range of values over which the fit allowed the parameters to vary. The model allows for up to three ISM absorption comments; we have simplified this table to }
\label{tab:MgIIFitParameters}
\end{deluxetable}

We modified the \cite{Youngblood2022} ISM model to include up to three absorption components. 
The optical depth, $\tau$, of each absorption component is described by a Voigt profile that depends on 
the column density ($N$(\MgII)), Doppler broadening parameter ($b_{MgII}$), and radial velocity ($V_{MgII}$) of the ISM cloud the light passes through. Each ISM absorption component, $F_{ISM}$ is defined as:

\begin{equation}
    F_{ISM} = exp(-\tau(N(MgII),b_{MgII},V_{MgII})).
\end{equation}

 \noindent When less than three ISM components were needed, we adopted $F_{ISM}$ = 1 for the unnecessary components.  
 
 The final model that we fit to our data is $F_{observed}$, the product of $F_{intrinsic}$ and three $F_{ISM}$ components, which is then convolved (represented by the symbol $\circledast$) with the instrument line spread function (LSF) provided by STScI\footnote{https://www.stsci.edu/hst/instrumentation/stis/performance/spectralresolution} for the appropriate grating and slit size: 

 \begin{equation}
    F_{observed} = (F_{intrinsic} \cdot F_{ISM,1} \cdot F_{ISM,2} \cdot F_{ISM,3}) \circledast LSF.
\end{equation}
 
The eighteen model parameters of $F_{observed}$, listed in Table \ref{tab:MgIIFitParameters}, were fit to the spectra using \texttt{lmfit}, a nonlinear least squares python fitting package. Our fitting code is provided in a publicly-available GitHub package, \texttt{MgII\_selfreversal}\footnote{https://github.com/allisony/MgII\_selfreversal}.

This model was fit to the spectrum for every star in our sample. Self-reversal parameters and velocities with the uncertainties returned from \texttt{lmfit} are reported in Table \ref{tab:refined}, and we show a subset of the fits in Figure \ref{fig:BestFitEx}. We validated our fit results by comparing our ISM parameter values against previously published values from \cite{Redfield2002}, \cite{2018ApJ...859...42Z}, \cite{2019ApJ...880..117E}, \cite{2021AJ....161..136C}, and \cite{Youngblood2022}. We generally found excellent agreement with published ISM parameter values, however our fit results sometimes indicated a different number of ISM components than found in the literature. In these cases, we employed the Bayesian Information Criterion (BIC) returned by \texttt{lmfit} and selected the best fit model with the lowest BIC value. When there were no literature results to compare with, we confirmed that \texttt{lmfit} returned a reduced chi-squared as close to $1.0$ as feasible before finalizing our result.The median $\chi^2_{\nu}$ of our sample was $0.8$ with 75\% of the $\chi^2_{\nu}$ values lying between 0.6 and 1.4. 
Some stars have large $\chi^2_{\nu}$ values between $2.0$ and $6.0$ and HD 128621 ($\alpha$ Cen B) has an extreme $\chi^2_{\nu}$ value of $47.3$. We find that these large $\chi^2_{\nu}$ values occur for bright stars with extremely high signal-to-noise spectra; this reveals that there are small features in the \MgII\ k line that are not included by our model and improvements could be made.

The fit's ability to accurately recover the self-reversal value is impacted by the presence of ISM absorption and the signal-to-noise (S/N) of the spectrum. In cases where the ISM absorption is coincident with the stellar emission line core, the uncertainty in the self-reversal parameter found by \texttt{lmfit} does not always accurately reflect the true uncertainty in $p$. Therefore, we assigned a fit quality number (1 being the worst to 3 being the best) to quantify our confidence in the self-reversal value obtained. We visually inspected each fit for the spectrum quality, location of the ISM absorption, and error bars on the fit parameters. In cases where the ISM absorption overlapped with the line core significantly (i.e., $V_{MgII} \approx V_{\rm radial}$) we found that \texttt{lmfit} did not appropriately inflate the uncertainty on $p$ to reflect the large degeneracy between deep self-reversal and deep ISM absorption. We therefore assigned a fit quality number of 1 to these spectra; this was often uncorrelated with S/N of the spectrum. There were also a few stars with no ISM contamination yet extremely low S/N to which we also assigned a fit quality number of 1.  The difference between a fit quality number 2 and 3 was determined by the S/N and ISM absorption location and width. The difference in quality factors are exemplified in Figure \ref{fig:BestFitEx} where the left-most column represents the lowest fit quality and the right-most column represents the highest fit quality. Generally, the M stars and distant F stars have lower S/N spectra resulting in larger uncertainties on the self-reversal parameter. These stars were only assigned the lowest quality factor if it appeared that the parameter uncertainties determined by the fit seemed unrealistically low based on the S/N and location of the ISM absorption.  Overall, we labeled 41 stars with a fit quality number of 1, 55 stars with a 2, and 40 stars with a 3.

\subsection{Mg II Self-Reversal Trends}

\begin{figure}
    \centering
    \includegraphics[width=0.8\textwidth]{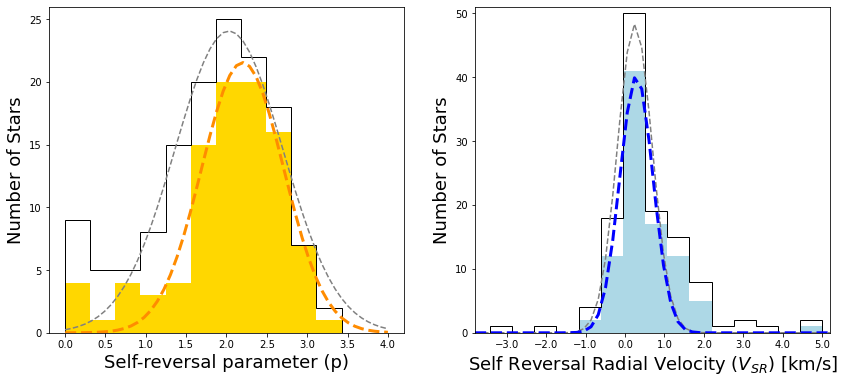}
    \caption{The distribution of self-reversal depths (left) and self-reversal radial velocities (right) are shown. The white histograms with black outlining show distributions for the full sample, while the colored histograms show distributions after excluding stars with the lowest quality parameter value. Dashed gray lines show Gaussian fits to the full sample distribution and dashed colored lines show Gaussian fits to the high-quality sample.} 
    \label{fig:Double_Histogram}
\end{figure}

The majority of stars in our sample exhibit self-reversal. This is evidenced by the histograms of self-reversal depths shown in Figure \ref{fig:Double_Histogram}. The distribution of values after excluding stars with quality parameter values $<$2 roughly follows a normal distribution with mean $p$=2.2$\pm$0.05 and standard deviation of 0.51$\pm$0.05, but there is an excess of stars with $p<$1. 
We calculated the  Shapiro-Wilk test for normality for each Gaussian fit in Figure \ref{fig:Double_Histogram} utilizing the Python package \texttt{scipy.stats.shapiro} package \citep{shapstat} including the test statistic ($s$) and the false alarm probability ($f$). We use a false-alarm probability threshold of $f$ $<$ 0.01 to identify statistically significant effects. For the entire sample of self-reversal parameter values we find $s$ = 0.96 with $f = 3.8 \times 10^{-4}$, and for the self-reversal parameter sample limited to a quality parameter $>$1 we find  $s$ = 0.93 with $f = 1.4 \times 10^{-4}$.  For the entire sample of self-reversal radial velocity values we find $s$ = 0.82 with $f = 4.90 \times 10^{-9}$, and for the self-reversal radial velocity sample limited to a quality parameter $>$1 we find  $s$ = 0.87 with $f = 4.75 \times 10^{-9}$. The distribution of the self-reversal parameter and the self-reversal velocity are closely Gaussian, based on the goodness of fit statistics (s $\sim$ 1) and the low false-alarm probabilities ($f$ $<$ 0.01). The mean value shows that the majority of stars in our sample exhibit less self-reversal than the Sun ($p$=2.59$\pm$0.04; \citealt{Youngblood2022}). We also examined the distribution of self-reversal radial velocities and compared the distribution to the solar value (0.79$\pm$0.07 \kms; \citealt{Youngblood2022}) in Figure \ref{fig:Double_Histogram}. The self-reversal radial velocity is a free parameter in our model that describes the asymmetry of the self-reversal within the \MgII\ emission line; it represents an offset from the fitted stellar radial velocity. Excluding stars with quality parameter values $<$2 and assuming a normal distribution, we find a mean of 0.29$\pm$0.05 \kms\ and standard deviation of 0.79$\pm$0.07 \kms, which is consistent with the solar value. There is an excess of stars with self-reversal radial velocities $>$1 \kms, in excess of the solar value. Positive radial velocities manifest in the spectra as blue asymmetries, meaning a brighter blue (shorter wavelength) peak compared to the red (longer wavelength) peak. Asymmetries indicate velocity gradients in the stellar atmosphere, although determining an absolute direction of flow is not straightforward \citep{Athay1970b,Cram1972,Linsky1980}.

\section{Analysis}\label{sec:results}

\begin{figure}[t!]
    \centering
    \includegraphics[scale=0.45]{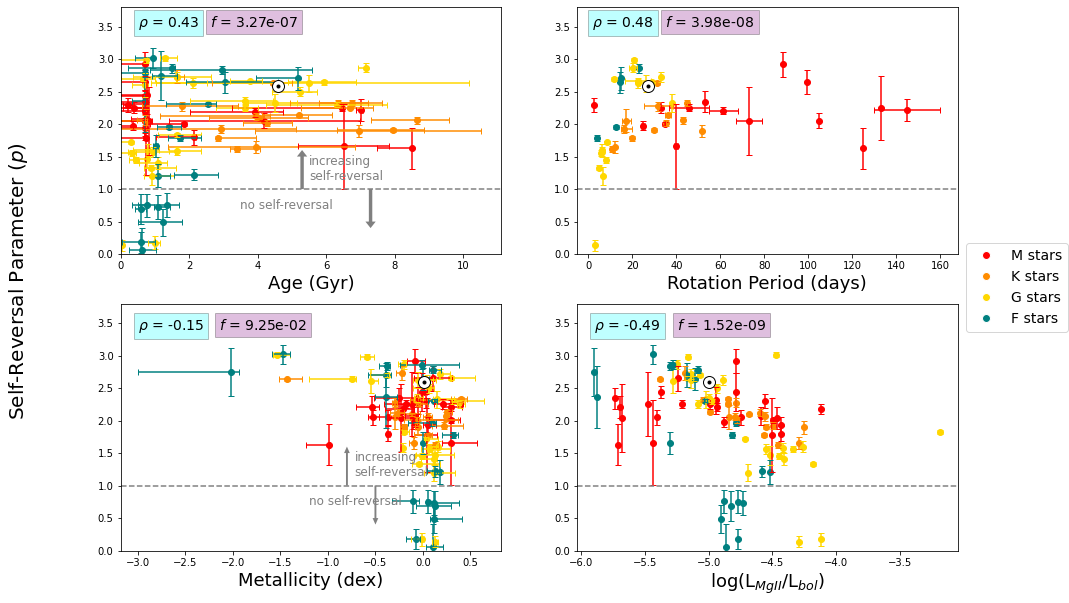}
    \caption{The self-reversal parameter versus stellar age (\textit{top left}), rotation period (\textit{top right}), metallicity (\textit{bottom left}), and the logarithm of the L$_{MgII}$/L$_{bolometric}$ luminosity ratio (\textit{bottom right}). The data points are color-coded by spectral type; M stars are shown in red, K stars in orange, G stars in yellow, and F stars in teal. Target stars without literature-established ages or rotation periods are not included in the top panels, respectively. Since the values of L$_{MgII}$ are integrals of the best-fit emission line of each star, we do not calculate uncertainties. Note that only 81 of our \ntargets\ target stars have measured rotation periods, and of those, 55 have high-quality fits that we show here.}
    \label{fig:mag}
\end{figure}

\begin{figure}[t!]
    \centering
    \includegraphics[width=1.0\textwidth]{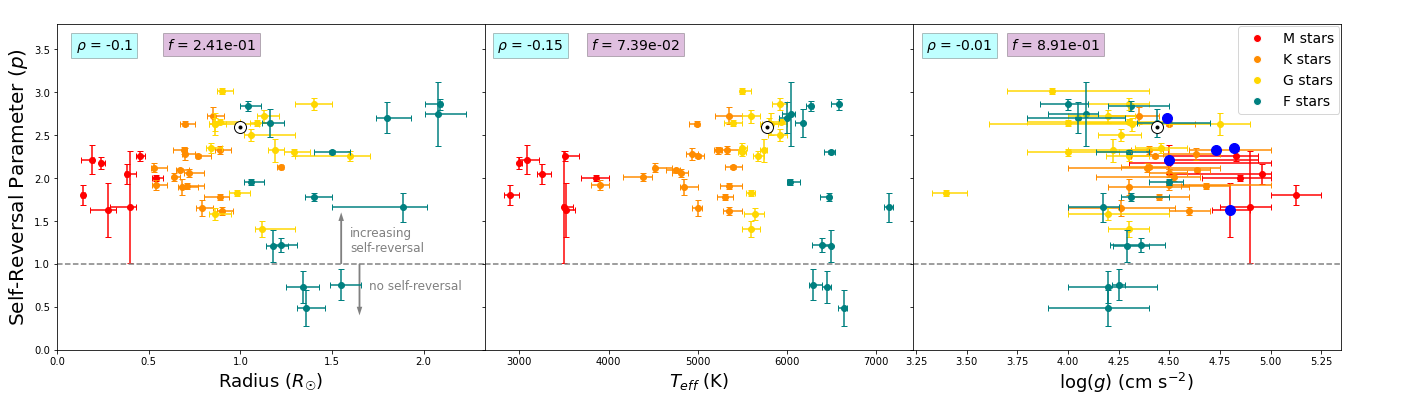}
    \caption{The self-reversal parameter versus radius (\textit{left}), \teff (\textit{middle}), and surface gravity (\textit{right}) for our field-age ($\geq$ 1 Gyr) target stars. In the right plot, the stars used in \cite{Youngblood2022} are highlighted in blue over our target stars and the Sun is represented by the solar symbol.}
    \label{fig:spect}
\end{figure}

In Figures \ref{fig:mag} and \ref{fig:spect}, we show comparisons between the \MgII\ self-reversal parameter and various intrinsic stellar parameters. We excluded any target stars with a fit quality number of 1 in our analysis in order to prevent any biases from low-quality \MgII\ self-reversal fits. The target stars with a fit quality of 2 and 3 are not differentiated on our plots since they are distributed similarly for each stellar parameter.


We utilize the Spearman rank-order correlation coefficient ($\rho$) and false alarm probability ($f$) calculated using the Python package \texttt{scipy.stats.spearmanr} to examine correlations between the self-reversal parameter and intrinsic stellar parameters. We use a false-alarm probability threshold of $f$ $<$ 0.01 to identify statistically significant effects. For the parameters shown in Figure \ref{fig:mag}, we find similar correlation coefficients and statistically significant false alarm probabilities for age ($\rho$= 0.43, $f$ =  $3.27 \times 10^{-7}$) and rotation period ($\rho$ = 0.48, $f$ = $3.98 \times 10^{-8}$). Although our determined age values do not agree with literature-cited ages as well as other parameters, we find that when comparing literature-cited ages and self-reversal, we find a similar correlation with $\rho$= 0.39, $f$ =  $1.61 \times 10^{-9}$. Note that, after a literature search, we find only 81 of our \ntargets\ target stars have measured rotation periods, and of those, only 25 had reported uncertainties. We find no significant correlations between self-reversal parameter and radius, effective temperature, surface gravity, or metallicity (Figures~\ref{fig:mag} and \ref{fig:spect}). We do find similar correlation coefficients and false alarm probabilities for effective temperature ($\rho$= -0.15, $f$ = $7.4 \times 10^{-2}$) and radius ($\rho$ = -0.10, $f$ = $2.69 \times 10^{-1}$). Since age and rotation period correlate for main-sequence stars \citep{Wilson1963, 1972Skumanich}, and mass and temperature also correlate for main-sequence stars \citep{1923Hertzsprung}, it follows that these pairs of stellar parameters would show similar correlation coefficients with the self-reversal parameter for our sample of main-sequence stars. We confirm that our target stars have correlated ages and rotation periods  ($\rho$ = 0.51, $f$ = $2.89 \times 10^{-6}$) as well as radii and temperatures ($\rho$ = 0.90, $f$ = $3.33 \times 10^{-52}$).

In Figure \ref{fig:spect}, we compare the self-reversal parameter and radius, $T_{\rm eff}$, and \logg, focusing on field age ($\geq 1$ Gyr) stars only in order to remove overlapping effects of age on potential correlations. We find that as radius and $T_{\rm eff}$  increase, the mean self-reversal depth stays roughly constant ($p$ = 1.97, 2.06, 2.37, and 1.93 for M, K, G, and F stars, respectively; i.e., no correlation as noted previously) but the variation in the self-reversal parameter values steadily increases from Ms to Gs (standard deviation $\sigma$ = 0.23, 0.28, 0.42, 0.80 for M, K, G, and F stars, respectively). 
This highlights differences between the behavior of the self-reversal depth for each spectral type, suggesting that cooler field-age stars (M, K) present with more similar self-reversal depths ($p$ = 1 - 2.7) while hotter field-age stars (G, F) present a broader range of self-reversal parameters ($p$ = 0.5 - 3). Although we attempted to avoid any interfering age effects in these relations, the increased spread with earlier spectral types appears to be due to age. Among the FGK stars, the low $p$ values in Figure \ref{fig:spect} correspond to stars with ages between 1-3 Gyr, while the higher p values are older stars. In our sample, there are no M or K dwarfs of any age that do not exhibit self-reversal, whereas one young G dwarf (0.04 Gyr) and several old and young F dwarfs (0.3 - 2 Gyr) lacked self-reversal. Contrary to \cite{Aryes1979}, in which they found that the self-reversal depth of the \CaII\ K line, another optically thick chromospheric emission line, negatively correlates with surface gravity, we find no clear correlation between the self-reversal parameter and \logg, consistent with the results from the smaller sample size used in \cite{Youngblood2022}. 

\begin{figure}[h!]
    \centering
    \includegraphics[scale=0.45]{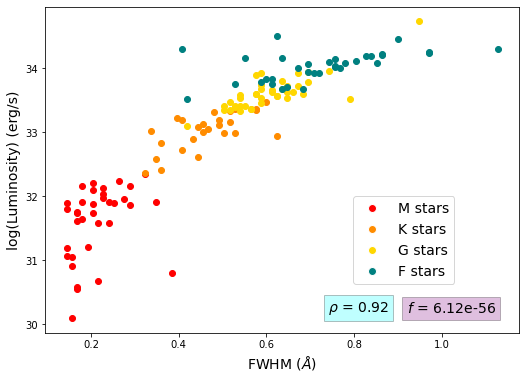}
    \caption{The bolometric luminosity versus the \MgII\ full width at half maximum. The data points are color-coded by spectral type.} 
    \label{fig:fhwm}
\end{figure}

We have considered whether this lack of correlation between self-reversal and \logg\ or other parameters like temperature and radius could be due to the limitation we imposed on \logg\ in our sample. Section~\ref{sec:obs} describes how we excluded very young or evolved stars because of the added complication that absorption from their winds would add to the analysis. To explore the possibility that our sample size range is too small to detect real, but scattered correlations that would be apparent over larger parameter ranges, we searched for the Wilson-Bappu effect \citep{WilsonBappu1957} in our sample. The Wilson-Bappu effect is a positive correlation between chromospheric line width and absolute stellar magnitude that has been observed over 10 orders of magnitude in bolometric stellar luminosity and for multiple chromospheric emission lines. We computed the \MgII\ Wison-Bappu line widths following \cite{Cassatella2001} and compared to the logarithm of the bolometric luminosity, which was determined using the Stefan-Boltzmann law. The logarithm of the bolometric luminosity is proportional to the absolute magnitude and more straightforward to compute given the uniform availability of effective temperature and radius for our sample. We detect the Wilson-Bappu correlation at high significance (Figure~\ref{fig:fhwm}); $\rho$ = 0.92, $f$ = 6.12$\times$10$^{-56}$; reported for all stars, regardless of whether we include the lowest quality fits), indicating that our sample size is sufficient to measure correlations between self-reversal and fundamental stellar parameters like $T_{\rm eff}$, radius, and \logg, if they existed. 
Note that we do not detect a significant correlation between the Wilson-Bappu \MgII\ line width and self-reversal parameter ($\rho$ = -0.22, $f$ = 0.03).

We find that younger stars with fast rotation periods are less likely to present with deep reversals. Youth and fast rotation periods are typically indicative of higher levels of magnetic activity \citep{Wilson1963, Wilson1966...144..695W}, and so, the identified correlation between age and P$_{rot}$ suggests that magnetic activity plays a role in determining the self-reversal depth in Mg II k, with higher activity stars being more likely to have Mg II k in full emission. These results show that stars with rotation periods greater than $\sim$10 days tend not to have a self-reversal parameter of less than 1.5, and stars with an age greater than 2 Gyr tend not to have a self-reversal parameter of less than 1.  We find a significant negative correlation between the self-reversal parameter and the ratio of the \MgII\ luminosity normalized to bolometric luminosity ($L_{\rm{MgII}}$/$L_{\rm{bol}}$), with $\rho$=-0.49 and $f$=1.52$\times$10$^{-9}$. The \MgII\ flux, along with other chromospheric emission lines (e.g., \CaII\ H \& K) are indicators of surface magnetism \citep{doyle1987activity, 1972Skumanich, 1992A&A...266..347P, 2017ARA&A..55..159L} and larger luminosity ratios indicate higher activity levels. Since we find that age, rotation period, and the \MgII\ luminosity ratio correlate with the self-reversal parameter, all three of which have been found to be magnetic activity indicators, our findings support that magnetic activity plays a role in determining the depth of the \MgII\ k self-reversal. This conclusion is further supported by the findings of \cite{1992A&A...266..347P} that showed a negative correlation between proxies for stellar activity (the intensity of the line core) and the self-reversal depth (the peak intensity of the line divided by the intensity of the line core) for the \CaII\ K line.  

\section{Summary}\label{sec:con}


 We have examined correlations between intrinsic stellar parameters and the chromospheric Mg II self-reversal depth using newly refined and consistently determined stellar parameters and Mg II line profiles for a sample of \ntargets\ stars. To generate a set of consistently determined stellar parameters, we computed a suite of 1D stellar upper atmosphere models with prescriptions for the chromosphere and transition region. To quantify the self-reversal depth, we simultaneously fit a model of the \MgII\ emission line with up to three ISM absorption components. 
 
 In modeling the photosphere and upper atmospheres of our main-sequence target stars, we found that for FGKM stars, visible and near-infrared photometry is significantly more sensitive to small (e.g., \teff\ $\pm$ 50 K, \logg\ $\pm$ 0.25 dex, \z\ $\pm$ 0.2 dex) changes in intrinsic stellar parameters compared to large changes in the chromospheric structure (e.g., $\nabla m_{Tmin} \pm\ 100\ g/cm^2$, $\nabla m_{TR} \pm\ 100\ g/cm^2$, $\nabla \Delta T_{TR}\ \pm$ 100 K). We therefore use a single low-activity upper atmospheric structure for all our stellar models.

Self-reversal ($p\ge$1) was present in $\sim$90\% of stars in our sample. 
We have identified no statistically significant correlations between self-reversal depth and any of the intrinsic stellar parameters, however there are some broad trends that we summarize here. We find that the cooler, less massive stars (K and M spectral type) all exhibit self-reversal, regardless of age or activity. The small number of stars that do not exhibit self-reversal are all young ($<$ 2 Gyr) F or G dwarfs. The range of observed self-reversal depths for each spectral type steadily increases from Ms to Fs, with the age of the star becoming a more influential factor controlling the depth of the self-reversal as we move towards earlier spectral types. We find that young stars with fast rotation periods may present with or without self-reversals, but older stars exclusively present with reversals. Finally, among G and K dwarfs, larger L$_{MgII}$/L$_{bol}$ values clearly correspond to shallower self-reversals; this distribution with L$_{MgII}$/L$_{bol}$ is roughly flat for the M dwarfs and has large scatter for the F dwarfs. Collectively, these results suggest that magnetic activity plays a role in determining the depth of a star's self-reversal, although is this may not be the only factor, for example, convection-driven wave heating may also contribute (e.g. \cite{1995ApJ...440L..29C}).

\section*{Acknowledgments}
This research was supported by HST-GO-15190.006-A, HST-GO-16701.001, and the Southeastern Universities Research Association. The contribution provided by S.P. is supported by NASA under award number 80GSFC21M0002.

\software{Astropy \citep{Robitaille2013}, IPython \citep{Perez2007}, Matplotlib \citep{Hunter2007}, NumPy and SciPy \citep{VanderWalt2011}, NumPyro \citep{phan2019composable,bingham2019pyro}, Pandas \citep{mckinney-proc-scipy-2010}, LMFIT \citep{newville_matthew_2014_11813}.} 

\bibliography{refs}{}
\bibliographystyle{aasjournal}

\end{document}